\documentclass[lettersize,journal]{IEEEtran}
\UseRawInputEncoding
\usepackage{amsmath,amsfonts}
\usepackage{algorithmic}
\usepackage{algorithm}
\usepackage{array}
\usepackage[caption=false,font=normalsize,labelfont=sf,textfont=sf]{subfig}
\usepackage{textcomp}
\usepackage{stfloats}
\usepackage{url}
\usepackage{verbatim}
\usepackage{graphicx}
\usepackage{cite}
\usepackage{multirow}
\usepackage{booktabs}
\usepackage{array}
\usepackage{caption}
\usepackage{makecell}
\usepackage[margin=1in]{geometry}
\usepackage{subcaption}
\usepackage{subfig}
\usepackage[table]{xcolor}
\definecolor{mygray}{RGB}{245, 245, 245} 
\usepackage{tabularx}
\usepackage{siunitx}

\begin{document}

\title{A Survey on LLM-based News Recommender Systems}

\author{Rongyao Wang, Veronica Liesaputra, Zhiyi Huang ~\IEEEmembership{Staff,~IEEE,}
\thanks{This paper was produced by the IEEE Publication Technology Group. They are in Piscataway, NJ.}
\thanks{Manuscript received April 19, 2021; revised August 16, 2021.}}

\markboth{Journal of \LaTeX\ Class Files,~Vol.~14, No.~8, August~2021}%
{Shell \MakeLowercase{\textit{et al.}}: A Sample Article Using IEEEtran.cls for IEEE Journals}


\maketitle

\begin{abstract}
News recommender systems play a critical role in mitigating the information overload problem.
In recent years, due to the successful applications of large language model technologies, researchers have utilized Discriminative Large Language Models (DLLMs) or Generative Large Language Models (GLLMs) to improve the performance of news recommender systems.  
Although several recent surveys review significant challenges for deep learning-based news recommender systems, such as fairness, privacy-preserving, and responsibility, there is a lack of a systematic survey on Large Language Model (LLM)-based news recommender systems.
In order to review different core methodologies and explore potential issues systematically, we categorize DLLM-based and GLLM-based news recommender systems under the umbrella of LLM-based news recommender systems.
In this survey, we first overview the development of deep learning-based news recommender systems. 
Then, we review LLM-based news recommender systems based on three aspects: news-oriented modeling, user-oriented modeling, and prediction-oriented modeling.
Next, we examine the challenges from various perspectives, including datasets, benchmarking tools, and methodologies. 
Furthermore, we conduct extensive experiments to analyze how large language model technologies affect the performance of different news recommender systems. 
Finally, we comprehensively explore the future directions for LLM-based news recommendations in the era of LLMs.     

\end{abstract}


\begin{figure}[htbp]
	\centering
	\includegraphics[width=0.5\textwidth]{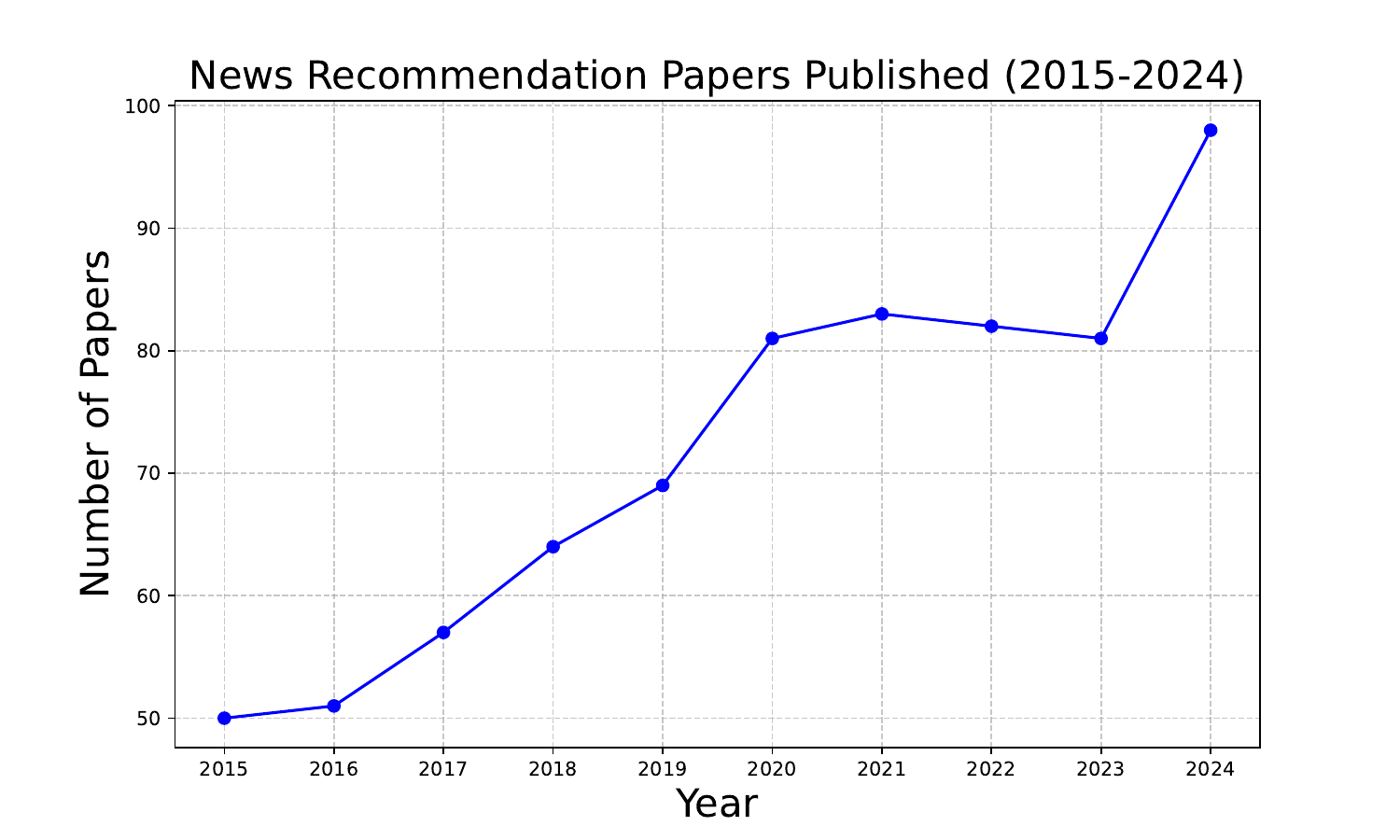}
	\caption{The tendency of news recommendation papers published between 2015 and 2024, based on Google Scholar data.}
	\label{fig:surveyfig}
\end{figure}

\section{Introduction}
\IEEEPARstart{A} large amount of news is generated in the era of Internet. It is difficult for users to find news that they are interested in from online news applications such as Google News, Bing News, and Toutiao. 
In these applications, news recommender systems are employed to help users alleviate this information overload as well as to improve users' reading experience \cite{wu2019npa, zheng2018drn}. 
In recent years, increasing attention to news recommendation has led to a growing number of publications on news recommender systems as shown in Figure \ref{fig:surveyfig}.
With the growth of deep-learning techniques in Natural Language Processing (NLP), various deep-learning methods are utilized in the news recommender systems and achieve state-of-the-art performances\cite{wu2023personalized}.
 
Most news recommender systems are built on deep neural network frameworks, such as Convolutional Neural Networks (CNNs), Recurrent Neural Networks (RNNs), and Graph Neural Networks (GNNs), due to their superior abilities at representing and learning textual information. 
CNNs are leveraged to extract local textual features from news contents \cite{wang2018dkn,wang2020fine,wu2019neurala}, such as fine-grained news features, and knowledge-based news features.
RNNs are utilized to capture users' diverse preferences based on their behavior sequences \cite{An2019LSTUR,wang2022news}, for example, long- and short-term user interest and multiple user interest. Furthermore, GNNs are commonly employed to model structural news-user representations \cite{hu2020graph,mao2021neural,qiu2022graph,wang2023intention} such as structural intent-aware user representation. Although these neural networks can improve the performance of the news recommender systems, many researchers have found that these general deep-learning methods tend to reach their limitations at learning more complex information \cite{vaswani2017attention,wu2021fastformer,wu2021empowering,10.1145/3695461} such as deep news-user relationship and multi-modal representation.

With the advancement of Transformers \cite{devlin2019bert,wolf2020transformers,wu2024survey} in NLP, researchers utilize DLLMs, such as BERT \cite{devlin2019bert}, RoBERTa \cite{liu2019roberta} and BART \cite{lewis2019bart}, as news encoders to capture the potential semantic information in news content \cite{wu2021empowering,bi2022mtrec,li2022miner,mao2023unitrec}, or apply DLLMs' special training strategy to model news-user semantic relationship for news recommendation \cite{zhang2021amm, wu2021newsbert, zhang2021unbert}.
These DLLM-based methods achieve better performance than deep neural networks (\textit{e.g.}, CNNs, RNNs, GNNs).
However, DLLM-based news recommender systems are constrained by their limited pre-trained knowledge, which makes it challenging to leverage them effectively for addressing cold-start problem and modeling accurate news and users' representations \cite{liu2019roberta, devlin2019bert, wu2021newsbert}. 

Recent GLLMs (\textit{e.g.}, GPT-4 \footnote{https://openai.com/index/gpt-4/}, LLaMA \footnote{https://www.llama.com/}, PaLM \footnote{https://ai.google/discover/palm2}) have a substantially larger number of parameters and are pre-trained on significantly higher amount of data, which makes it more powerful at semantic understanding and generation. 
Recently, there are rapid growth in GLLM-based news recommender systems, and some can achieve state-of-the-art performance \cite{gao2024generative, ruan2024rewriting, chen2024lkpnr} because it can alleviate the cold-start problem by generating relevant information, and use its strong reasoning and learning abilities to explore accurate news features and model users' interests.
However, GLLM-based news recommender systems typically require significant training time and resources.

Many comprehensive survey studies summarize and review various methodologies of news recommender systems. 
For instance, Wu et al. \cite{wu2023personalized} reviewed different challenges, technologies, and future directions of deep learning based personalized news recommender systems. 
While Meng et al. \cite{meng2023survey} reviewed various significant parts of personalized news recommender system such as data collection, news recommendation model, and personalized news display. 
They focused on discussing different news recommendation methods based on graph structure learning. 
Iana et al. \cite{iana2024survey} categorized knowledge-aware news recommendation into three parts: neural methods, non-neural entity-centric methods, and non-neural path-based methods. 
Specifically, they review different methodologies of news recommender systems which utilize external knowledge to enhance performance.
However, currently, there is a gap in a systematic survey on LLM-based news recommender systems with extensive experiments.

In this paper, we summarize and review various LLM-based (including DLLM and GLLM) news recommendation approaches based on three aspects: news-oriented modeling, user-oriented modeling, and prediction-oriented modeling. 
Text-oriented modeling is defined as a set of foremost text-based encoding methods that use DLLMs and GLLMs for news recommendation. 
Then, we review methods that focus on building user profiles as user-oriented modeling for news recommendation. 
The prediction-oriented modeling represents methods relevant to prediction, \textit{i.e.}, optimizing predicting function based on users and news. 
Moreover, we examine the challenges from various perspectives, including datasets, benchmarking tools, and methodologies.
We have also conducted comprehensive benchmark experiments to thoroughly compare and review the existing systems' performance, advantages, disadvantages, and limitations.
The performance of each news recommender system is measured in terms of classification metrics, ranking metrics, diversity metrics, and personalization metrics.
Finally, we review future directions on new recommendation in the era of LLMs to support future news recommendation research.

Our main contributions are as follows:
\begin{itemize}
	\item As far as we know, this is the first systematic survey of news recommender systems by conducting extensive experiments in the era of LLMs. 
	\item We propose a unified research framework that reviews different LLM-based news recommendation models based on three aspects: news-oriented modeling, user-oriented modeling, and prediction-oriented modeling.
	\item Through objective experiments, we evaluate various LLM-based news recommender systems from different angles using a variety of metrics, \textit{i.e.}, classification metrics, ranking metrics, diversity metrics, and personalization metrics.  

\end{itemize}

This paper is organized as follows:
Section \ref{section2} discusses different deep learning-based news recommendation methodologies.
While, Section \ref{section3} reviews LLM-based news recommendation methods from three aspects: news-oriented modeling, user-oriented modeling, and prediction-oriented modeling.
Section \ref{section4} explores the challenges of LLM-based news recommender systems.
Section \ref{section5} details our benchmark experiments' designs and results of various LLM-based news recommendation methodologies.
Section \ref{section6} explores future directions for LLM-based news recommender systems.
Finally, in Section \ref{section7}, we conclude our systematic survey and significant findings.

\section{Overview of deep learning-based news recommender systems}
\label{section2}


\begin{figure}[htbp]
	\centering
	\includegraphics[width=0.5\textwidth]{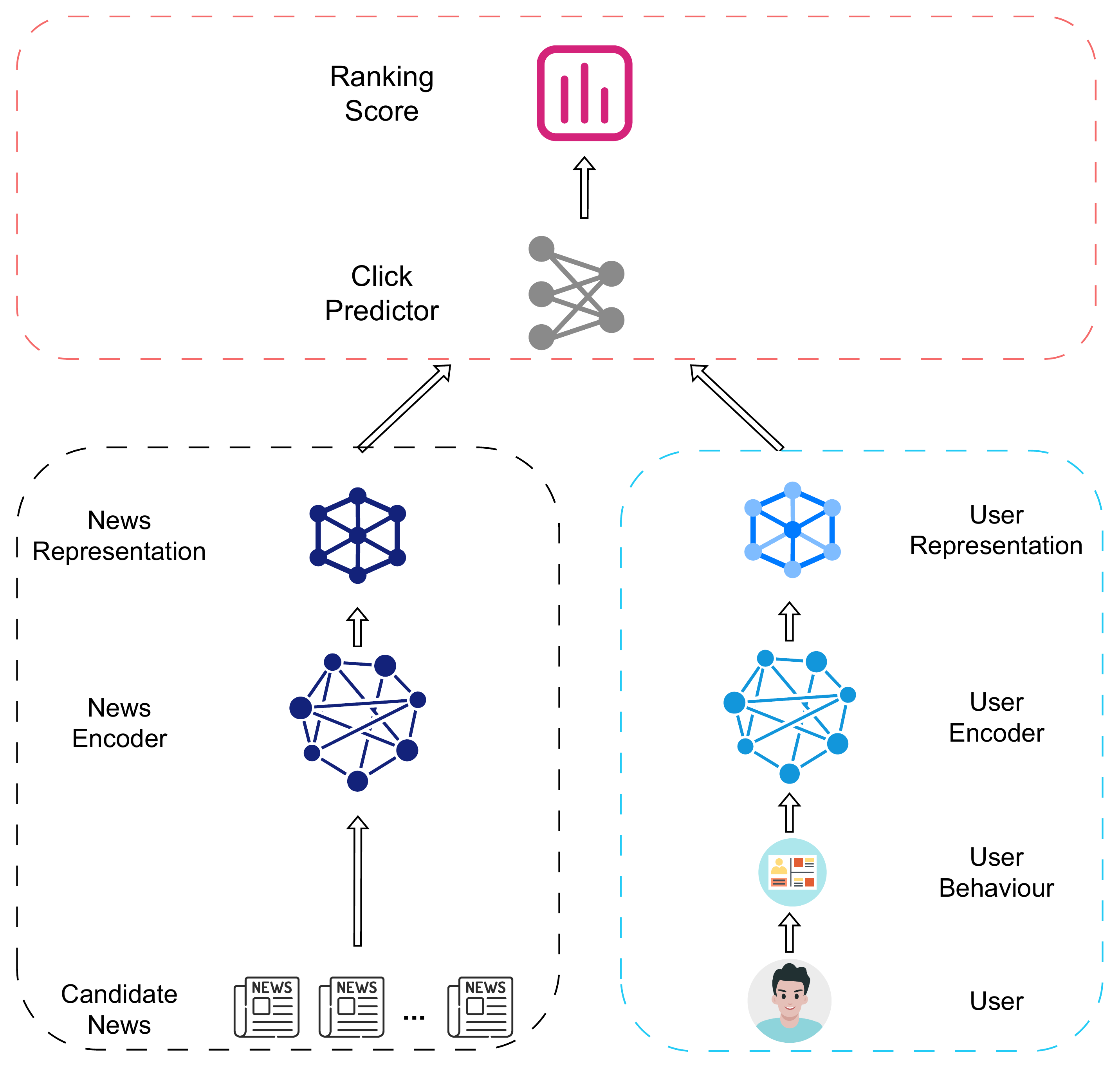}
	\caption{A general uniform news recommendation framework}
	\label{fig:image1}
\end{figure}

Before the era of large language models, deep-learning methods (\textit{e.g.}, CNNs, RNNs, GNNs) were commonly used to design news recommender systems \cite{zhu2019dan, shi2021wg4rec, khattar2018hram, hu2021trnews, okura2017embedding, pang2020time, lee2020news}.
Research works in each period have demonstrated consistent patterns across different deep-learning technology development stages.
Based on our knowledge and previous works \cite{wu2021empowering, iana2024survey, wu2023personalized}, we can conclude a general uniform news recommendation framework as shown in Figure \ref{fig:image1}, which consists of three main components: news encoder, user encoder and click predictor.
Specifically, the news encoder is devised to construct accurate news representations and support user interest modeling.
The user encoder is designed to explore users' preferences and build users' representations based on accurate news representations.
The click predictor is proposed to calculate matching scores between candidate news representations and users' representations.
In this section, we categorize different research works into two types based on the characteristics of news recommendation as follows: news-oriented modeling and user-oriented modeling.
To clearly illustrate our observations, an overview of deep learning-based news recommendation methodologies is presented in Table \ref{tab:summary}. 

\begin{table*}[htbp]
  \centering
  \caption{Overview of deep learning-based news recommender systems in recent years. We categorize the research based on their methodologies.}
  \begin{tabular}{lp{5cm}p{7cm}} 
    \toprule
    \textbf{Type} & \textbf{Methodology} & \textbf{Model} \\
    \midrule
    \multirow{5}{*}{News-oriented modeling} 
      & Convolutional Neural Network (CNN) & Weave\&Rec \cite{khattar2018weave}, DNN4NR \cite{park2017deep} \\
      & Recurrent Neural Network (RNN) & GRU \cite{okura2017embedding}, CHAMELEON \cite{de2018chameleon}, HRAM \cite{khattar2018hram} \\
      & Attention Mechanism & NRMS \cite{wu2019nrms}, DAN \cite{zhu2019dan}, DAINN \cite{zhang2019dynamic}, NPA \cite{wu2019npa}, DNA \cite{dna2019dynamic} \\
      & Graph Neural Network (GNN) & SI-News \cite{zhu2022si}, GERL \cite{ge2020graph} \\
      & Knowledge Graph (KG) & DKN \cite{wang2018dkn}, TEKGR \cite{lee2020news}, KOPRA \cite{tian2021joint}, AnchorKG \cite{liu2021reinforced}, KRED \cite{liu2020kred}, CAGE \cite{sheu2021knowledge} \\
      & Reinforcement Learning & DRN\cite{zheng2018drn} \\
      & News Representation Learning & TANR \cite{wu2019neurala}, NAML \cite{ijcai2019-536}, WG4Rec \cite{shi2021wg4rec}, ANRS\cite{lu2022aspect} \\
    \midrule
    \multirow{5}{*}{User-oriented modeling} 
    & User Representation Learning & LSTUR \cite{An2019LSTUR}, HiFi-Ark\cite{liu2019hi}, NRHUB\cite{wu2019nrhub}, CNE-SUE\cite{mao2021neural}, DFM\cite{lian2018towards}, TANN\cite{pang2020time}, GNewsRec\cite{hu2020graph}, HieRec\cite{qi2021hie}, PP-Rec\cite{qi2021pp}, CAUM\cite{qicaum2022}\\
      & Interest Modeling & MINS \cite{wang2022news}, MINER\cite{li2022miner}, FIM\cite{wang2020fine}, PENR\cite{wang2021popularity}, GREP\cite{qiu2022graph}, TrNews\cite{hu2021trnews} \\
      & Intention Modeling & IPNR \cite{wang2020intention} \\
      & Sentiment Modeling & SentiRec \cite{wu2020sentirec}, SentiDebias\cite{wu2022removing} \\
      & Fairness Modeling & FairRec\cite{wu2021fairness}, HDInt\cite{wang2024hierarchical} \\
      & Privacy-preserving Modeling & CenNewsRec \cite{qi2020privacy}, FedRec\cite{yi2021efficient} \\
      & Satisfaction Modeling & CPRS\cite{wu2020cprs} \\
 
    \bottomrule
  \end{tabular}
  \label{tab:summary}
\end{table*}

\subsection{News-oriented Modeling}
Initially, most researchers focus on applying basic deep-learning methods to construct news embeddings.
To be specific, CNNs are among the earliest deep-learning methods applied in news recommender systems.
For example, Wang et al. \cite{wang2018dkn} managed to design CNN-based news encoders with a knowledge graph.
Khattar et al. \cite{khattar2018weave} utilized 3D-CNNs to encode news representations.
Moreover, attention mechanism (\textit{e.g.}, additive attention, multi-head self-attention) is employed to model accurate news representation \cite{wu2019nrms, zhu2019dan} as well as capture textual content representations \cite{wu2019npa}.
In particular, Gabriel et al. \cite{de2018chameleon} proposed a deep learning-based news recommendation architecture that applied deep neural networks (DNNs) and RNNs to encode content representations.
Wu et al. \cite{wu2019neurala} designed a topic-aware news encoder to build news representations with news categories.
To construct comprehensive news representations, other researchers focus on devising effective news encoders to encode informative news features.
To capture sufficient textual information in news content, Wu et al. \cite{ijcai2019-536} proposed a multi-view news encoder that can encode all informative news features such as title, category, and abstract.
Lian et al. \cite{lian2018towards} proposed applying multi-layer fully connected networks and attention mechanisms to design a deep fusion model for news representation learning.
The aforementioned models encode basic news features (\textit{e.g.}, title, category, abstract) with different deep-learning methods.
Due to the limitations of deep-learning methods during this period, most researchers are interested in exploring different deep-learning methodologies to improve the accuracy of news recommender systems.

\subsection{User-oriented Modeling}
With the growth of deep learning-based news recommender systems, user representation learning has gained increasing attention from researchers.
For example, Liu et al. \cite{liu2019hi} utilized self-attention networks to encode deep user representations.
Aiming to capture users' long- and short-term representations, An et al. \cite{An2019LSTUR} proposed to employ RNNs (\textit{e.g.}, gated recurrent unit networks) to model users' consistent and temporal preferences.
Ge et al. \cite{ge2020graph} employed GNNs (\textit{e.g.}, graph attention networks) and modified transformer networks to model users' high-order relatedness.
Although these representation learning methods are proposed to encode news and users' features in order to generate accurate embedding representations for news recommendation, they are only at the mid-exploration stage of deep learning-based news recommender systems.
  
During the rapid development stage of deep-learning methods ranging from 2020 to 2023, researchers have observed that deep-learning methods not only explore more accurate textual features but also model real-world user profiles (\textit{e.g.}, interest, intention, behaviour).
Aiming to build accurate users' interests and obtain more effective recommendation performance, most works utilize different sufficient deep-learning methods to devise their innovative user interest models \cite{wang2020fine, qi2021hie, qi2021pp, wang2021popularity, qiu2022graph}.
They all aim to extract diverse user interests from historical records and achieve state-of-the-art performance in their experiments.
By leveraging effective and accessible deep learning methods, we can obtain accurate representations of user interests as expected.
For instance, applying RNNs could help our model capture long sequential features \cite{wang2022news}, while GNNs could assist in modeling structural interest representations and high-order relationships between news and users \cite{mao2021neural}.
Moreover, user intention is a critical concept in news recommendation, originating from the field of psychology \cite{wang2020intention}.
In order to model users' intentions from sequential history records, Wang et al. \cite{wang2023intention} devised a GNN-based framework to extract intentions with a knowledge graph.
In addition to modeling user representations, other aspects of the user experience (\textit{e.g.}, privacy, fairness, sentiment) should also be taken into account in news recommender systems.
In terms of privacy, Yi et al. \cite{yi2021efficient} proposed using federated learning to protect users' privacy while reducing computation and communication costs. 
For fairness, Wu et al. \cite{wu2021fairness} first employed decomposed adversarial learning to learn bias-free representations in the news recommender system. 
Furthermore, Wu et al. \cite{wu2020sentirec} devised to model sentiment-aware news representations with a sentiment prediction task.
The aforementioned research directions have become prominent research topics in the field of news recommender systems in recent years.

\section{LLM-based news recommendation methodology} \label{LMs-based news recommendation framework}
\label{section3}

In this section, we review how LLM-based methods are used to construct the three main components of a news recommendation system: news-oriented modeling, user-oriented modeling, and prediction-oriented modeling. 

\subsection{News-oriented Modeling}
Initially, most researchers utilize various DLLMs to encode the textual news content.
Specifically, Wu et al. \cite{wu2021empowering} first proposed the use of BERT to enhance the natural language understanding capabilities of news recommender systems.
Their core contribution is replacing the conventional news encoder module consisting of CNNs, RNNs and GNNs with a module composed of DLLMs and attention mechanisms.
Huang et al. \cite{huang2023personal} introduced a multi-factor fusion model to address the impact of specific news events (\textit{i.e.}, breaking news). 
To enable multi-aspect customization for news recommendation, such as incorporating news features like sentiment and categories, Iana et al. \cite{iana2023train} proposed a novel framework that leverages DLLMs and contrastive learning to model both content- and aspect-based news representations. 
These methods commonly leverage BERT-based models to enhance the comprehension of textual information for news recommendation.
As demonstrated in these studies\cite{iana2023newsreclib, wu2021empowering}, BERT-based models have enhanced news recommender systems, achieving approximately a 3\% improvement in performance compared to deep learning-based methods.

The application of GLLMs in news recommender systems is distinct from the approach used in DLLM-based news recommender systems.
At the inception of GLLMs, most researchers favored incorporating additional critical information generated by GLLMs into news recommender systems.
For example, Gao et al. \cite{gao2024generative} proposed a novel generative news recommendation framework that constructs news narratives (\textit{e.g.}, personalized multi-news narratives) using GLLMs, with a newly proposed training method to enhance recommendation accuracy.
Similarly, Yada et al. \cite{yada2024news} applied GLLMs (\textit{i.e.}, GPT-4) to generate category descriptions in the news recommender system.
Chen et al. \cite{chen2024lkpnr} effectively leveraged GLLMs to construct rich semantic news representations and utilize Wikidata KG to tackle the long-tail problem in news recommender systems.
Liu et al. \cite{liu2024once} proposed a novel framework to apply large language model technolegies (\textit{e.g.}, GPT-4, LLaMA) to learn contextual information with effective prompts and fine-tuning methods on the news recommendation dataset (\textit{e.g.}, MIND). 
In order to alleviate the problem of media bias, Ruan et al. \cite{ruan2024rewriting} proposed leveraging GLLMs to rewrite news headlines for users in news recommender systems.
The researchers effectively leverage the powerful generative capabilities of GLLMs to achieve approximately a 10\% improvement in the performance of news recommendation. 

\subsection{User-oriented Modeling}
Modeling users' preferences lies at the heart of news recommender systems.
An advanced modeling method could enhance news recommendation models to understand users' needs and recommend relevant articles that meet their preferences.
The applications of LLMs in news recommender systems not only improve the ability to understand textual information but also enable the creation of accurate user profiles by leveraging contextual features.
For instance, Zhang et al. \cite{zhang2021unbert} designed a user-news matching framework with BERT, whose core idea is matching clicked and candidate news with multi-grained representations (\textit{i.e.}, word-level representations, news-level representations) in a BERT architecture.
During the same period, the attentive multi-field matching framework is proposed by Zhang et al. \cite{zhang2021amm}.
Although their methodology is similar to that of  \cite{zhang2021unbert}, the matching objects and granularity differ (\textit{i.e.}, news titles, abstracts, and bodies are used in this matching process).
Similarly, Li et al.\cite{li2022miner} proposed a multi-interest matching framework that applies a poly attention scheme to extract multiple interests with BERT-based news encoders.
To effectively model users' immediate and long-term preferences, Liu et al. \cite{liu2024modeling} utilized attention mechanisms and GLLMs to address these challenges, using clicked news articles as the basis.
As far as we observed, GLLM-based user models are less than DLLM-based user models.
Therefore, a significant gap exists in modeling user preferences using GLLMs.

\subsection{Prediction-oriented modeling}
The prediction-oriented modeling consists of a combination of training strategies and prediction methods.
Specifically, NewsBERT \cite{wu2021newsbert} is a novel DLLM-based news recommendation framework that simultaneously learns valuable insights from both teacher and student models through the distillation of BERT.
Yu et al. \cite{yu2022tiny} first proposed a self-supervised domain-specific post-training approach into a DLLM-based news recommendation framework with a novel two-stage knowledge distillation methodology. 
Moreover, Xiao et al. \cite{xiao2022training}  proposed an innovative training framework SpeedyFeed to reduce the time and resource costs associated with training DLLM-based news recommender systems.
To enhance the performance of news recommendation tasks, Bi et al. \cite{bi2022mtrec} designed a multi-task framework that improves effectiveness by incorporating multi-field features, including news recommendation, news classification, and named entity recognition (NER).
Furthermore, Liu et al. \cite{liu2023recprompt} proposed a prompt-based news recommendation framework that demonstrates the effectiveness of a GLLM-based prompt strategy by leveraging an iterative bootstrapping process.
Wang et al. \cite{wang2024cherryrec} proposed a GLLM-based news recommendation model that utilizes a GLLM to filter out low-value news and recommend high-value news to users, using a newly designed metric called CherryRec.
These methods provide great strategies to assist us train and optimize LLM-based news recommender systems.
However, research on GLLM-based news recommender systems is still in the early stages and presents researchers with more opportunities and challenges to overcome.

\section{Challenges}
\label{section4}
In this section, we discuss our observations on the challenges of LLM-based news recommender systems, focusing on datasets, tools, and methodologies.

\begin{table*}[htbp]
	\centering
	\caption{Comparisons of different news recommendation datasets}
	\begin{tabular}{cccccp{2cm}cp{3cm}}
		\hline
		Released Year & Dataset & \# News & \# User & \# Behavior & Textual Feature & Language & \multicolumn{1}{c}{Reference} \\
        \hline
		2013 & Plista \footnote{https://dai-labor.de/en/publications/the-plista-dataset-2/}  & 70,353  & /         & 1,095,323    & title, body, category                & German         & Jugovac et al. \cite{jugovac2018streamingrec} \\
		\hline
		2018 & Globo \footnote{https://www.kaggle.com/datasets/gspmoreira/news-portal-user-interactions-by-globocom}   & 46,000  & 314,000   & 3,000,000    & word embeddings of texts   & Portuguese     & Gabriel et al.\cite{de2018chameleon} \\
		\hline
		2018 & Adressa \footnote{https://reclab.idi.ntnu.no/dataset/} & 48,486  & 3,083,438 & 27,223,576   & title, body, category, entity & Norwegian   & Dan et al. \cite{zhu2019dan}, Hu et al. \cite{hu2020graph}, Sheu et al. \cite{sheu2021knowledge}, Koo et al. \cite{koo2020accurate}, zhao et al. \cite{zhao2024dynamic}, Gharahighehi et al. \cite{gharahighehi2021fair}, Bae et al. \cite{bae2023lancer}, Huang et al. \cite{huang2023personal}, Yi et al. \cite{yi2021efficient}, Lee et al. \cite{lee2020news} \\
		\hline
		2020 & MIND \footnote{https://msnews.github.io/}    & 161,013 & 1,000,000 & 24,155,470   & title, abstract, category, entity & English   & Wang et al. \cite{wang2018dkn}, Wu et al. \cite{ijcai2019-536, wu2019neurala, wu2019npa, wu2019nrms, wu2020sentirec, wu2021empowering, wu2021fairness, wu2021fastformer, wu2021newsbert}, An et al. \cite{An2019LSTUR}, Wang et al. \cite{wang2020fine}, Qi et al. \cite{qi2021hie, qi2021pp, qicaum2022}, Qiu et al. \cite{qiu2022graph}, Li et al. \cite{li2022miner}, Mao et al. \cite{mao2021neural, mao2023unitrec}, Wang et al. \cite{wang2022news, wang2023intention}, Zhang et al. \cite{zhang2021amm, zhang2021unbert}, Gao et al. \cite{gao2024generative}, Yi et al. \cite{yi2021efficient}, Ge et al. \cite{ge2020graph}, Liu et al. \cite{liu2019hi}, Wang et al. \cite{wang2021popularity}, Lu et al. \cite{lu2022aspect}, Bi et al. \cite{bi2022mtrec}   \\
		\hline
		2023 & NPR \footnote{https://www.kaggle.com/datasets/joelpl/news-portal-recommendations-npr-by-globo/data}   & 148,099 & 1,162,802 & 1,402,576   & title, content, topic            & Portuguese  & - \\
		\hline
		2024 & xMIND \footnote{https://github.com/andreeaiana/xMIND}   & 130,379 & 1,000,000 & 24,155,470   & title, abstract            & Multi-languages & - \\
		\hline
        2024 & EB-NeRD \footnote{https://recsys.eb.dk}   & 125,541 & 1,103,602 & 37,966,985   & title, abstract, body, category, subcategory, entities, URL, sentiment, topic            & Danish & - \\
		\hline
	\end{tabular}%
	\label{tab:table1}%
\end{table*}%

\subsection{Challenges of Datasets}
There are some publicly available news recommendation datasets such as Globo \cite{de2018news}, Plista \cite{kille2013plista}, Adressa \cite{gulla2017adressa}, MIND \cite{wu2020mind}, NPR \cite{lucas2023npr}, xMIND \cite{iana2024mind}, and EB-NeRD \cite{kruse2024eb}. 
We summarize their characteristics in Table \ref{tab:table1}.
Specifically, Plista offers a large collection of news texts and user behaviors gathered from 13 German news portals.
Globo only includes word embeddings of the news texts, without additional textual features, which significantly limits the available information.
Because textual information is essential in constructing news and user representations.
As we can observe, Adressa and MIND are two of popular news recommendation datasets, which include informative news and user features.
NPR is an enhanced news dataset by Globo, offering more comprehensive news content and detailed user behavior data, which consists of metadata information about news articles, recommendation impressions and user consumption history. 
xMIND is a multilingual news recommendation dataset built on MIND, which includes 14 linguistically and geographically diverse languages.
EB-NeRD is collected from the information of Ekstra Bladet, which consists of 37 million impression logs, 251 million interactions and news metadata.

Despite the availability of some resources, public news recommendation datasets face significant challenges as follows:
\begin{enumerate}
    \item Quantity: The available public news datasets are insufficient to meet the rapid growth of news recommender systems. 
As we can see in Table \ref{tab:table1}, MIND dataset is the most popular dataset on which most researchers prefer to conduct experiments.
However, the MIND dataset has nearly reached its limitation in some works \cite{wu2021empowering, wu2021newsbert}. 
We are in the zone of model overfitting to this dataset. 
So we need more datasets to really evaluate the generalizability of the model.
It is essential to release some new news datasets to enable researchers to conveniently design and conduct experiments across different datasets.
\item Information: MIND dataset contains the most informative features (\textit{e.g.}, title, abstract, category, entity) compared with others.
However, this alone is insufficient to significantly enhance news recommender systems. 
This is because additional information (\textit{e.g.}, publisher, location, reading time, etc) is required to model more accurate user representations.
Incorporating users' reading time per news article could improve the accuracy of modeling their preferences \cite{wu2020cprs}. 
Including publisher information could contribute to building fairness-aware news recommender systems \cite{wu2024survey}. 
Additionally, utilizing images within news articles could enable researchers to develop multi-modal news recommender systems, which can empower news representation learning \cite{wu2022mm}.
As a result, the development of news recommender systems is constrained by available news datasets.
Moreover, an LLM trained solely on English is unlikely to perform well on Norwegian or other language datasets without additional training.
\end{enumerate}

\begin{table*}[htbp]
	\centering
	\caption{Comparison of important benchmarking tools}
	\begin{tabular}{lp{1.5cm}lp{2cm}cp{1cm}p{1.5cm}p{1cm}}
		\hline
		Name & Program Framework & Program Language & \multicolumn{1}{c}{URL} & Author & \multicolumn{1}{c}{Year} & \multicolumn{1}{c}{Model} & \multicolumn{1}{c}{Dataset} \\
		\hline
		Recommenders & TensorFlow & Python & \url{https://github.com/recommenders-team/recommenders} & Microsoft & 2019 & DKN\cite{wang2018dkn}, NAML\cite{ijcai2019-536}, NPA\cite{wu2019npa}, LSTUR\cite{An2019LSTUR}, NRMS\cite{wu2019nrms} & MIND \\
		\hline
		RecBole & PyTorch & Python & \url{https://recbole.io/} & RUCAIBox & 2020 & NRMS\cite{wu2019nrms}, NAML\cite{ijcai2019-536} & MIND \\
		\hline
		News-Recommendation & PyTorch & Python & \url{https://github.com/yusanshi/news-recommendation} & Yusanshi & 2019 & DKN\cite{wang2018dkn}, NAML\cite{ijcai2019-536}, NPA\cite{wu2019npa}, LSTUR\cite{An2019LSTUR}, NRMS\cite{wu2019nrms}, HiFi-Ark\cite{liu2019hi} & MIND \\
		\hline
		NewsRecLib & PyTorch-Lightening & Python & \url{https://github.com/andreeaiana/newsreclib} & Iana et al.\cite{iana2023newsreclib} & 2023 & CAUM\cite{qicaum2022}, CenNewsRec\cite{qi2020privacy}, DKN\cite{wang2018dkn}, LSTUR\cite{An2019LSTUR}, MINER\cite{li2022miner}, MINS\cite{wang2022news}, NAML\cite{ijcai2019-536}, NPA\cite{wu2019npa}, NRMS\cite{wu2019nrms}, TANR\cite{wu2019neurala}, MANNeR\cite{iana2023train}, SentiDebias\cite{wu2022removing}, SentiRec\cite{wu2020sentirec} & MIND, Adressa, xMIND \\
		\hline
		Legommender & PyTorch & Python & \url{https://github.com/Jyonn/Legommenders} & Liu et al\cite{legommenders}. & 2024 & NAML\cite{ijcai2019-536}, LSTUR\cite{An2019LSTUR}, NRMS\cite{wu2019nrms}, PLMNR\cite{wu2021empowering},ONCE\cite{liu2024once} & MIND \\
        \hline
	\end{tabular}%
	\label{tab:tools}%
\end{table*}

\subsection{Challenges of Benchmarking Tools}
Smart and user-friendly benchmarking tools enable researchers to conduct studies more easily.
In recent years, several benchmarking tools have been released to support the development of news recommender systems.
We collect and review important benchmarking tools related to news recommendation illustrated in Table \ref{tab:tools}, including Microsoft Recommenders \cite{graham2019microsoft}, RecBole \cite{recbole}, News-Recommendation, NewsRecLib \cite{iana2023newsreclib}, and Legommender \cite{legommenders}.

\subsubsection{Microsoft Recommenders}
Microsoft Recommenders \footnote{https://github.com/recommenders-team/recommenders} is published by Microsoft Recommenders team providing several examples and practices for building recommender systems.
This benchmarking tool contains five essential news recommendation models including NRMS \cite{wu2019nrms}, NPA \cite{wu2019npa}, NAML \cite{ijcai2019-536}, LSTUR \cite{An2019LSTUR} and DKN \cite{wang2018dkn}, which code is generated by  Keras \footnote{https://keras.io/} in Tensorflow \footnote{https://github.com/tensorflow/tensorflow} programming environment.
Although this repository offers several news recommendation models for practical use, it is not a specialized benchmarking tool for news recommendation. and only conducts experiments on a single news recommendation dataset (\textit{i.e.}, MIND).

\subsubsection{RecBole}
RecBole \footnote{https://recbole.io/} is a general recommendation benchmarking tool developed by PyTorch \footnote{https://pytorch.org/}, which contains various recommendation algorithms such as sequential recommendation, context-aware recommendation, and knowledge-based recommendation.
This repository also supports news recommender systems, but only two models (NRMS \cite{wu2019nrms} and NAML \cite{ijcai2019-536}).
Therefore, RecBole is insufficient for researchers studying news recommendation due to its limited number of models and reliance on a single news recommendation dataset.

\subsubsection{News-Recommendation}
News-Recommendation \footnote{https://github.com/yusanshi/news-recommendation} is a personal open-source benchmarking tool available on GitHub.
This benchmarking tool consists of six important news recommendation models such as DKN \cite{wang2018dkn}, NAML\cite{ijcai2019-536}, NPA\cite{wu2019npa}, NRMS\cite{wu2019nrms}, TANR\cite{wu2019neurala} and HiFi-Ark\cite{liu2019hi}.
Additionally, the author uses MIND datasets to conduct experiments.
Although the repository implements various news recommendation models, the author will likely discontinue updating the code.
As a result, these models are outdated and unable to support future research.

\subsubsection{NewsRecLib}
NewsRecLib\footnote{https://github.com/andreeaiana/newsreclib} is a news recommendation benchmarking tool built by a personal researcher based on PyTorch Lightning \footnote{https://lightning.ai/docs/pytorch/stable/} and Hydra \footnote{https://hydra.cc/} frameworks.
This benchmarking tool contains various deep learning-based news recommender systems (\textit{e.g.}, TANR \cite{wu2019neurala}, CAUM \cite{qicaum2022}, CenNewsRec \cite{qi2020privacy}, MINS \cite{wang2022news}, etc) and news datasets (\textit{e.g.}, MIND, Adressa, xMIND), which enables researchers to quickly reproduce news recommendation methods and conduct experiments among different datasets.
Although this benchmarking tool categorizes news recommender systems into two classes: deep learning-based news recommendation and fairness-aware news recommendation, the models provided are insufficient to support future research due to the lack of multi-modal news recommender systems, GLLM-based news recommender. systems, etc.
Several future directions require attention, including debiased news recommender systems, LLM-based news recommender systems, multi-modal news recommender systems, and privacy-preserving news recommender systems.

\subsubsection{Legommender}
Legommender is a content-based recommendation benchmarking tool released by personal researchers, which aims to support LLM-based recommender systems.
This benchmarking tool contains four news recommendation models such as NAML\cite{ijcai2019-536}, NRMS\cite{wu2019nrms}, LSTUR\cite{An2019LSTUR}, and PLMNR\cite{wu2021empowering}.
They devised this tool to serve their research such as ONCE\cite{liu2024once}, SPAR\cite{zhang2024spar}, GreenRec\cite{liu2024benchmarking} and UIST\cite{liu2024discrete}.
Legommender guides researchers on effectively applying LLMs to news recommendations including DLLM- and GLLM-based frameworks.
However, it is not sufficient to support a wide range of news recommendation models with LLMs.
Furthermore, additional news recommendation datasets are expected to better serve LLM-based news recommender systems.

\subsection{Challenges of Methodologies}
LLM-based news recommender systems face several challenges in the era of LLMs.
We will review these various challenges from three aspects: news-oriented modeling, user-oriented modeling, and prediction-oriented modeling.

\subsubsection{News-oriented Modeling}
News representation modeling is critical for accurate news recommendation, strongly related to NLP technologies such as word embedding\cite{okura2017embedding} and LLMs.
LLM-based news recommender systems utilize LLMs as news encoders to learn informative news representations, and their superior performance has been validated in numerous studies \cite{wu2021empowering, liu2024once, iana2023newsreclib}.
Although language models offer promising advantages for news representations, challenges still need to be addressed.
First of all, the information generated by GLLMs is not always credible (\textit{i.e.}, GLLMs may exhibit hallucinations.) \cite{ravichander2025halogen}.
Therefore, it is essential to develop more reliable methodologies to validate the accuracy of generated representations.
Second, processing large volumes of textual information with LLMs consumes significant time and resources \cite{yin2025enhancing}.
Third, LLM-based news recommender systems are unable to process long documents due to their limited context window, which restricts their ability to construct comprehensive news representations.
Furthermore, modeling multiple languages poses a significant challenge for LLM-based news recommender systems, as not all LLMs are equipped to handle multiple languages effectively.
In summary, more effective and efficient LLM-based news encoders are needed to better leverage LLMs for constructing informative news representations.

\subsubsection{User-oriented Modeling}
User-oriented modeling plays a critical role in news recommender systems.
As far as we know, most researchers prefer to model user and news representations simultaneously using DLLMs \cite{zhang2021unbert, zhang2021amm, wu2021newsbert, iana2023train, huang2023personal, li2022miner, yu2022tiny}.
As a result, there is limited research focused on building special user representations independently such as multiple interests modeling, intention modeling, and sentiment modeling \cite{wang2022news, wang2023intention, wu2020sentirec}.
In the real world, user behaviors are complex and dynamic.
Current user representation models cannot describe complex and dynamic user behaviors.
For example, users' interests might be influenced by breaking news along with time.
It is essential to carefully analyze and understand users' interests based on their behaviors.
Although DLLMs enhance news representations as well as user representations, it is not sufficient to fully understand real complex users' behaviors.
On the other hand, limited research has focused on modeling user representations using GLLMs \cite{gao2024generative, chen2024lkpnr}.
They propose generating additional information such as narratives by GLLMs and then aggregating it with the original features of news articles.
However, these published studies ignore exploring deep and various user behaviours based on GLLM-enhanced context.
For example, user behaviours when reading online newspapers may be influenced by their purpose and preferences.
Therefore, efficiently building complex and dynamic user representations using GLLMs remains a significant challenge for news recommendation.

\subsubsection{Prediction-oriented Modeling}
Prediction-oriented modeling is related to training strategies and prediction including different training frameworks, evaluation methods, and ranking methodologies.
It is more and more pivotal for news recommender systems in the period of LLMs.
Due to the high time and resource consumption of DLLMs and GLLMs, LLM-based news recommender systems require optimization during training and prompting.
First of all, LLM-based news recommender systems often apply one-tower architecture to match candidate news and user interests \cite{zhang2021amm, zhang2021unbert}.
These ranking methods can accurately explore the relatedness between candidate news and user interests with multi-grained features.
However, these methods are unsuitable for low-latency or low-resource scenarios due to their high inference time and resource requirements \cite{wu2024survey}.
Moreover, training open-source GLLMs, such as LLaMA, for use in news recommendation requires significant time and expensive hardware to support these experiments.
Efficient and effective training frameworks and ranking methods are required to accelerate prediction without high time and resource consumption.
Second, current LLM-based methods lack accountability for their generated results.
Many researchers have already utilized LLMs to replace traditional prediction frameworks such as the two-tower model as shown in Figure \ref{fig:image1} \cite{zhang2023prompt,li2024prompt, wang2024multi, wang2023large}.
However, the generated results by GLLMs are not credible.
Therefore, how to evaluate the accuracy of the generated information is a significant challenge.
Moreover, new metrics are required to support LLM-based news recommender systems in order to verify their robustness, reliability, and performance.

\section{Experiments}
\label{section5}
In this section, we introduce experimental settings such as datasets, metrics, and news recommendation models that we used in our experiments.
Further, we conduct extensive experiments in order to answer the following questions:

\textbf{Q1}: How is the performance of LLM-based news recommendation models compared with deep learning-based news recommendation models in terms of classification and ranking? 

\textbf{Q2}: Do LLM-based news recommendation models outperform deep learning-based news recommendation models in terms of diversity and personalization?

\textbf{Q3}: Do LLMs improve the performance of news recommendation sharply compared with the original deep learning-based news recommendation models?

\textbf{Q4}: How do GLLM-based news recommendation models perform compared with DLLM-based news recommendation models?

\subsection{Datasets}
For our experiments, we will utilize the two most commonly used datasets in our experiments: MIND \cite{wu2020mind} and Adressa \cite{gulla2017adressa}.

\subsubsection{MIND}
MIND was released by Microsoft in 2020 \cite{wu2020mind} and constructed based on anonymous user logs obtained from Microsoft News platform.
There are two versions: MIND-small and MIND-large.
MIND-small contains 65,238 news metadata and 347,727 logs generated by 94,057 users. MIND-large consists of 24,155,470 logs of 1,000,000 users and 161,013 pieces of news.
Each piece of news contains titles, abstracts, categories, and entities。
In terms of users, MIND provides their IDs and clicked logs including news clicks and impressions.

\subsubsection{Adressa}
Adressa was published by Norwegian University of Science and Technology, which contains a collection of news articles and sessions from Adressavisen news platform \cite{gulla2017adressa}.
There are two sub-datasets: one-week and ten-week.
The one-week Adressa dataset is comprised of 11,207 articles and 2,286,835 session logs of 561,733 users.
The 10-week Adressa dataset is composed of 48,486 articles and 27,223,576 session logs of 3,083,438 users.
Each news article consists of titles, categories, bodies, and entities. 
Each session log includes various information about the user such as IDs, regions, time, browser, and city. 

Due to the limitation of training time and resources, we use the MIND-small and one-week Adressa dataset to conduct our experiments.

\subsection{Evaluation Metrics}
Several evaluation metrics are utilized to verify the performance of news recommender systems in terms of ranking, classification, diversity, and personalization.
In this section, we present the metrics employed in our experiments.

\subsubsection{Classification Metrics}
News recommendation can be considered as a classification task. 
Hence, several classification metrics can be used.
Area Under Curve (AUC) score is commonly used in the evaluation of news recommendations. 
A high AUC score indicates that the news recommendation model has a stronger ability to distinguish between negative and positive samples, enabling it to recommend relevant news to users effectively.
The score is calculated as follows:
\begin{equation}
\small
	\text{AUC} = \frac{1}{|P| \times |N|} \sum_{p \in P} \sum_{n \in N} I(s(p) > s(n)),
\end{equation}
where $P$ is positive samples, $N$ is negative samples, $p$ is one of positive samples, $n$ is one of negative samples, $s(p)$ is the predicting score of positive samples, $s(n)$ is the predicting score of negative samples.
$I(s(p) > s(n))$ indicates that if $s(p) > s(n)$, the result is 1; otherwise, it is 0, computed as follows:

\begin{equation}
\small
	I(s(p) > s(n)) = 
	\begin{cases} 
		1, & \text{if } s(p) > s(n) \\
		0, & \text{otherwise}
	\end{cases}.
\end{equation}
Additionally, there are other popular metrics used in the evaluation of news recommender systems such as Precision, Recall, and Hit Rate (HR), which are computed as follows:
\begin{equation}
\small
	\text{Precision} = \frac{TP}{TP + FP},
\end{equation}
\begin{equation}
\small
	\text{Recall} = \frac{TP}{TP + FN},
\end{equation}
\begin{equation}
\small
	\text{HR}@k = \frac{1}{|U|} \sum_{u \in U} \mathbb{I}(R_u \cap \hat{R}_u(k) \neq \emptyset),
\end{equation}
where $TP$ indicates true positive samples, $FP$ means false positive samples, $FN$ indicates false negative samples, $R_u$ is the collection of interested items for user $u$, $\hat{R}_u(k)$ is a top-$k$ list of interested items for user $u$, $\mathbb{I}$ is exponential function, if true, return 1; otherwise, return 0.
In summary, AUC is widely utilized to evaluate the models' ability to recognize positive and negative samples.
Precision measures the accuracy of correctly predicting positive samples.
Recall is widely used to verify the ability to distinguish all positive samples.
Hit rate is proposed to evaluate whether the recommended news list contains the user's interested news. 

\subsubsection{Ranking Metrics}
News recommendation can be considered as a ranking task.
Therefore, there are several ranking metrics used in the evaluation of news recommender systems.
Apart from AUC, which is also a ranking metric, Mean Reciprocal Rank (MRR) and Normalized Discounted Cumulative Gain (nDCG).
Specifically, MRR is utilized to evaluate whether interested news appears in the top positions of the recommended news list.
nDCG measures whether the most relevant news appears in the top positions of the recommended news list, which focuses on the ranking quality and relevance of recommended results.
MRR and nDCG\(@k\) are formulated as follows:
\begin{equation}
\small
	\text{MRR} = \frac{1}{|Q|} \sum_{i=1}^{|Q|} \frac{1}{\text{Rank}_i},
\end{equation}
\begin{equation}
\small
	\text{nDCG}@k = \frac{\sum_{i=1}^{k} \frac{\text{rel}_i}{\log_2(i + 1)}}{\sum_{i=1}^{k} \frac{\text{rel}^{*}_i}{\log_2(i + 1)}},
\end{equation}

where $Q$ is the set of queries, $|Q|$ is the total number of queries, $\text{Rank}_i$ is the rank position of the first relevant to the $i$-th query.
$\text{rel}_i$ is the relevance score of the $i$-th result, $\text{rel}^{*}_i$ is relevance score under ideal ranking.


\subsubsection{Diversity Metrics}
In order to measure the diversity of news recommendations, Iana et el.\cite{iana2023train} proposed aspect-based diversity metrics as follows:
\begin{equation}
\small
	{D_{{A_p}}}@k =  - \sum\limits_{j \in {A_p}} {\frac{{p\left( j \right)\log p\left( j \right)}}{{\log \left( {\left| {{A_p}} \right|} \right)}}},
\end{equation}
where $A_p$ is the collection of aspects, ${\left| {{A_p}} \right|}$ is the number of aspects.

\subsubsection{Personalization Metrics}
The Jaccard similarity \cite{bag2019efficient} is used to evaluate the personalization of news recommendations, which is computed as follows:
\begin{equation}
\small
	P{S_{{A_p}}}@k = \frac{{\sum\nolimits_{j = 1}^{\left| {{A_p}} \right|} {\min \left( {{{\cal R}_j},{{\cal H}_j}} \right)} }}{{\sum\nolimits_{j = 1}^{\left| {{A_p}} \right|} {\max \left( {{{\cal R}_j},{{\cal H}_j}} \right)} }},
\end{equation}
where $\cal H$ is the user history, $\cal R$ is the recommendation list, ${\cal H}_j$ and ${\cal R}_j$ are the probability of a piece of news with class $j$.
In our experiments, we denote $categ\_div$ as the topical category-based diversity and $sent\_div$ as sentiment-based diversity. 
Moreover, we consider $categ\_pers$ as the topical category-based personalization and $sent\_pers$ as sentiment-based personalization.

In our experiments, we compute top 5 and 10 scores in terms of nDCG, Hit, Recall, Precision, diversity, and personalization.

\subsection{Baselines}
In order to evaluate the performance of LLM-based news recommendation models, we conduct extensive experiments on various methodologies which are classified into three groups: deep learning-based news recommendation models, DLLM-based news recommendation models, and GLLM-based news recommendation.
\subsubsection{Deep Learning-based News Recommendation Models}
\begin{itemize}
	\item NPA \cite{wu2019npa}, a personalized news recommendation model, which applies users' IDs to improve the performance of news recommendation with CNNs and attention mechanisms.
	\item TANR \cite{wu2019neurala}, a topic-aware news recommendation model, which promotes news recommendations with a topic classification task using CNNs and attention mechanisms.
	\item NAML \cite{ijcai2019-536
	}, a deep learning-based news recommendation model with multi-viewing learning, which builds news representations from various news features such as titles, abstracts, and categories using CNNs and attention mechanisms.
	\item LSTUR \cite{An2019LSTUR}, a deep learning-based news recommendation model, which employs RNNs to learn long- and short-term user representations.
	\item NRMS \cite{wu2019nrms}, a deep learning-based news recommendation model, which utilizes multi-head self-attention mechanisms to capture complex news and users' features.
	\item CenNewsRec \cite{qi2020privacy
	}, a privacy-preserving news recommendation model, which applies federated learning to jointly train accurate news recommender systems on users' devices and servers.
	\item CAUM \cite{qicaum2022}, a candidate-aware news recommendation model, which learns candidate-aware user representations using self-attention mechanisms and CNNs in order to accurately match users' interests.
	\item MINS \cite{wang2022news}, a deep learning-based news recommendation model, which can learn multi-interest user representations through a multi-channel network consisting of RNNs and self-attention mechanisms.
	\item SentiDebias \cite{wu2022removing}, a deep learning-based news recommendation model, which employs decomposed adversarial learning to achieve fairness-aware news recommendation in terms of different sentiments.
	\item SentiRec \cite{wu2020sentirec}, a diversity-aware news recommendation model, which applies Transformers to model sentiment-aware news and user representations, jointly training with a sentiment prediction task. 
	    
\end{itemize}

\subsubsection{DLLM-based News Recommendation Models}
\begin{itemize}
	\item MINER \cite{li2022miner}, a DLLM-based news recommendation model, which proposes a poly attention scheme to learn multi-interest user representations over BERT.
	\item MANNeR \cite{iana2023train}, a DLLM-based news recommendation model, which proposes a modular framework to learn aspect-specific representations over BERT.
    \item LSTUR-DLLM, TANR-DLLM, NRMS-DLLM, and NAML-DLLM are DLLM-empowered news recommendation models, in which the original news encoders are modified to incorporate DLLMs, inspired by previous works \cite{wu2021empowering}, \cite{iana2023newsreclib}.
\end{itemize}

\subsubsection{GLLM-based News Recommendation Models}
\begin{itemize}
	\item LKPNR \cite{chen2024lkpnr}, a generative news recommendation framework, which integrates a knowledge graph and GLLMs into deep learning-based news recommendation models.  
    \item ONCE \cite{liu2024once}, a hybrid content-based recommendation framework, which leverages both open- and closed-source GLLMs to enhance the performance of content-based recommender systems including news recommender systems. 
\end{itemize}

\begin{table*}[htbp]
  \centering
  \renewcommand{\arraystretch}{1.5}
  \caption{This table illustrates the performance of different news recommendation approaches including deep learning-based news recommendation models and LLM-based news recommendation models in terms of classification and ranking metrics on the MIND and Adressa datasets. 
  Based on the analysis of these results, we can conclude that LLM-based news recommendation models outperform deep learning-based news recommendation models on the MIND dataset.
  The reported values represent the mean and standard deviation derived from five distinct experimental runs.
  Bold indicates the best experimental results on the same dataset. Underline means the second-best experimental results on the same dataset. Gary's color represents LLM-based news recommendation models.}
    \setlength{\tabcolsep}{7pt} 
    \begin{tabular}{c|c|cccccc}
    \toprule
    \multirow{2}[4]{*}{Dataset} & \multicolumn{1}{c|}{\multirow{2}[4]{*}{Model}} & \multicolumn{6}{c}{Metric} \\
    \cmidrule(lr){3-8}
          &       & AUC   & MRR   & NDCG@5 & NDCG@10 & Recall@5 & Recall@10 \\
    \midrule
    \multirow{14}[2]{*}{MIND} 
          & NPA   & 57.12$\pm$0.0035 & 30.61$\pm$0.0046 & 28.68$\pm$0.0036 & 34.89$\pm$0.0031 & 41.95$\pm$0.0043 & 59.98$\pm$0.0022 \\
          & LSTUR & 56.31$\pm$0.017 & 33.09$\pm$0.0057 & 31.31$\pm$0.0068 & 37.61$\pm$0.0068 & 45.37$\pm$0.0055 & 63.5$\pm$0.0044 \\
          & TANR  & 60.83$\pm$0.0066 & 32.88$\pm$0.0037 & 30.93$\pm$0.0045 & 37.19$\pm$0.0034 & 44.71$\pm$0.007 & 62.51$\pm$0.0047 \\
          & NRMS  & 55.63$\pm$0.0229 & 28.63$\pm$0.0158 & 26.78$\pm$0.014 & 33.3$\pm$0.0133 & 40.38$\pm$0.0135 & 59.17$\pm$0.0125 \\
          & NAML  & 50.17$\pm$0.0004 & \underline{34.55$\pm$0.0066} & 32.69$\pm$0.0055 & 38.93$\pm$0.0048 & 46.73$\pm$0.0048 & 64.59$\pm$0.0023 \\
          & CenNewsRec & 53.85$\pm$0.015 & 26.6$\pm$0.0105 & 24.98$\pm$0.0105 & 31.67$\pm$0.0092 & 39.05$\pm$0.0123 & 58.35$\pm$0.0078 \\
          & CAUM  & 60.63$\pm$0.0111 & 34.15$\pm$0.0091 & 32.23$\pm$0.0084 & 38.69$\pm$0.0076 & 44.75$\pm$0.0331 & 63.36$\pm$0.0294 \\
          & MINS  & 58.78$\pm$0.0161 & 33.76$\pm$0.0036 & 31.87$\pm$0.0034 & 38.29$\pm$0.0033 & 46$\pm$0.0066 & 64.39$\pm$0.0054 \\
          & SentiDebias & 54.82$\pm$0.0204 & 25.6$\pm$0.0142 & 23.3$\pm$0.0133 & 30.12$\pm$0.0129 & 35.59$\pm$0.0127 & 55.32$\pm$0.0105 \\
          & SentiRec & 52.87$\pm$0.0085 & 29.44$\pm$0.0188 & 27.18$\pm$0.0175 & 33.62$\pm$0.0167 & 40.25$\pm$0.0178 & 58.83$\pm$0.0159 \\
          & \cellcolor{mygray}LSTUR-DLLM & \cellcolor{mygray}50$\pm$0	& \cellcolor{mygray}30.15$\pm$0.0079 & \cellcolor{mygray}28.56$\pm$0.0077 & \cellcolor{mygray}34.92$\pm$0.0077 & \cellcolor{mygray}42.02$\pm$0.0112 & \cellcolor{mygray}60.27$\pm$0.0111 \\
          & \cellcolor{mygray}TANR-DLLM & \cellcolor{mygray}49.95$\pm$0.0013 & \cellcolor{mygray}25.32$\pm$0.0143 & \cellcolor{mygray}22.81$\pm$0.0148 & \cellcolor{mygray}29.17$\pm$0.0152 & \cellcolor{mygray}34.16$\pm$0.0222 & \cellcolor{mygray}52.61$\pm$0.0229 \\
          & \cellcolor{mygray}NRMS-DLLM & \cellcolor{mygray}50$\pm$0 & \cellcolor{mygray}20.77$\pm$0.0228 & \cellcolor{mygray}18.37$\pm$0.0231 & \cellcolor{mygray}24.83$\pm$0.0218 & \cellcolor{mygray}28.71$\pm$0.0327 & \cellcolor{mygray}47.63$\pm$0.0273 \\
          & \cellcolor{mygray}NAML-DLLM &\cellcolor{mygray}52.59$\pm$0.0189 & \cellcolor{mygray}30.13$\pm$0.0128 & \cellcolor{mygray}28.42$\pm$0.0123 & \cellcolor{mygray}34.95$\pm$0.012 & \cellcolor{mygray}41.89$\pm$0.0147 & \cellcolor{mygray}60.61$\pm$0.0146 \\
          & \cellcolor{mygray}MINER & \cellcolor{mygray}51.08$\pm$0.0062 & \cellcolor{mygray}24.9$\pm$0.0039 & \cellcolor{mygray}22.6$\pm$0.0062 & \cellcolor{mygray}28.85$\pm$0.0054 & \cellcolor{mygray}34.44$\pm$0.0083 & \cellcolor{mygray}52.58$\pm$0.0053 \\
          & \cellcolor{mygray}MANNeR & \cellcolor{mygray}\textbf{68.44$\pm$0.0088} & \cellcolor{mygray}\textbf{37.3$\pm$0.017} & \cellcolor{mygray}\textbf{35.67$\pm$0.0163} & \cellcolor{mygray}\textbf{41.79$\pm$0.0142} & \cellcolor{mygray}\underline{49.96$\pm$0.0134} & \cellcolor{mygray}\underline{67.4$\pm$0.0069} \\
          & \cellcolor{mygray}LKPNR & \cellcolor{mygray}\underline{67.32$\pm$0.0004} & \cellcolor{mygray}31.99$\pm$0.0003 & \cellcolor{mygray}\underline{35.46$\pm$0.0005} & \cellcolor{mygray}\underline{41.77$\pm$0.0002} & \cellcolor{mygray}\textbf{50.27$\pm$0.001} & \cellcolor{mygray}\textbf{68.21$\pm$0.0004} \\
          & \cellcolor{mygray}ONCE & \cellcolor{mygray}65.06$\pm$0.0031 & \cellcolor{mygray}32.76$\pm$0.0046 & \cellcolor{mygray}34$\pm$0.0054 & \cellcolor{mygray}40.17$\pm$0.0049 & \cellcolor{mygray}48.37$\pm$0.0054 & \cellcolor{mygray}65.9$\pm$0.0046 \\
 
    \midrule
    \multirow{12}[2]{*}{Adressa} 
          & NPA   & 52.72$\pm$0.0366 & 32.94$\pm$0.0223 & 32.51$\pm$0.0303 & 38.77$\pm$0.0387 & 45.93$\pm$0.0539 & 65.43$\pm$0.0879 \\
          & LSTUR & 68.37$\pm$0.0104 & \textbf{36.85$\pm$0.0194} & \textbf{37.68$\pm$0.029} & \textbf{44.85$\pm$0.0241} & \textbf{54.23$\pm$0.0462} & \underline{76.39$\pm$0.0439} \\
          & TANR  & 50.22$\pm$0.0028 & 33.99$\pm$0.0222 & 33.71$\pm$0.0323 & 41.5$\pm$0.0256 & 48.4$\pm$0.0492 & 72.65$\pm$0.0278 \\
          & NRMS  & 64.54$\pm$0.0259 & 30.35$\pm$0.0207 & 28.63$\pm$0.0285 & 38.85$\pm$0.0342 & 42.09$\pm$0.0444 & 73.78$\pm$0.0897 \\
          & NAML  & 50$\pm$0  & 35.52$\pm$0.016 & 35.15$\pm$0.0227 & 42.42$\pm$0.0213 & 48.99$\pm$0.0378 & 71.71$\pm$0.0499 \\
          & CenNewsRec & 64.77$\pm$0.0295 & 29.63$\pm$0.021 & 26.97$\pm$0.0305 & 36.74$\pm$0.0347 & 38.04$\pm$0.0533 & 68.62$\pm$0.0773 \\
          & CAUM  & \textbf{72.33$\pm$0.0475} & \underline{35.68$\pm$0.0327} & 36.04$\pm$0.0443 & \underline{44.42$\pm$0.0473} & \underline{52.48$\pm$0.0723} & \textbf{78.5$\pm$0.0778} \\
          & MINS  & 68.68$\pm$0.0486 & 33.49$\pm$0.039 & 32.13$\pm$0.0514 & 40.84$\pm$0.0489 & 45.04$\pm$0.0717 & 72.26$\pm$0.0796 \\
          & SentiDebias & \underline{68.7$\pm$0.0361} & 31.1$\pm$0.0361 & 30.51$\pm$0.0559 & 39.66$\pm$0.039 & 45.87$\pm$0.0868 & 74.19$\pm$0.0464 \\
          & SentiRec & 55.59$\pm$0.0096 & 27.14$\pm$0.0118 & 24.35$\pm$0.0107 & 30.78$\pm$0.0112 & 33.32$\pm$0.0082 & 53.49$\pm$0.0262 \\
          & \cellcolor{mygray}LSTUR-DLLM & \cellcolor{mygray}56.86$\pm$0.0334 & \cellcolor{mygray}34.34$\pm$0.039 & \cellcolor{mygray}35.58$\pm$0.0433 & \cellcolor{mygray}42.31$\pm$0.0509 & \cellcolor{mygray}48.66$\pm$0.0719 & \cellcolor{mygray}68.3$\pm$0.1049 \\ 
          & \cellcolor{mygray}TANR-DLLM & \cellcolor{mygray}50.03$\pm$0.0007 & \cellcolor{mygray}35.17$\pm$0.024 & \cellcolor{mygray}35.09$\pm$0.0277 & \cellcolor{mygray}41.27$\pm$0.0322 & \cellcolor{mygray}47.49$\pm$0.045 & \cellcolor{mygray}66.39$\pm$0.0669 \\
          & \cellcolor{mygray}NRMS-DLLM & \cellcolor{mygray}52.8$\pm$0.0516 & \cellcolor{mygray}28.89$\pm$0.1134 & \cellcolor{mygray}27.95$\pm$0.1382 & \cellcolor{mygray}33.48$\pm$0.1505 & \cellcolor{mygray}38.74$\pm$0.1898 & \cellcolor{mygray}55.23$\pm$0.2353 \\
          & \cellcolor{mygray}NAML-DLLM & \cellcolor{mygray}50$\pm$0.0001 & \cellcolor{mygray}33.43$\pm$0.0331 & \cellcolor{mygray}34.45$\pm$0.0314 & \cellcolor{mygray}40.65$\pm$0.0297 & \cellcolor{mygray}43.23$\pm$0.0593 & \cellcolor{mygray}68.64$\pm$0.0429 \\
          & \cellcolor{mygray}MINER & \cellcolor{mygray}56.96$\pm$0.0741 & \cellcolor{mygray}26.5$\pm$0.1134 & \cellcolor{mygray}25.68$\pm$0.1253 & \cellcolor{mygray}32.21$\pm$0.1256 & \cellcolor{mygray}32.28$\pm$0.2281 & \cellcolor{mygray}66.7$\pm$0.1282 \\
          & \cellcolor{mygray}MANNeR & \cellcolor{mygray}50$\pm$0  & \cellcolor{mygray}35.35$\pm$0.0489 & \cellcolor{mygray}\underline{36.38$\pm$0.0687} & \cellcolor{mygray}44.19$\pm$0.0602 & \cellcolor{mygray}51.11$\pm$0.0979 & \cellcolor{mygray}75.66$\pm$0.0766 \\
    \bottomrule
    \end{tabular}%
  \label{tab:4}%
\end{table*}

\subsection{Experimental Setup}  
We use similar configurations with NewsRecLib\cite{iana2023newsreclib} in terms of deep learning-based news recommendation models and DLLM-based news recommendation models.
Besides, we employ official codes to reproduce GLLM-based news recommendation models (\textit{e.g.}, ONCE\footnote{https://github.com/Jyonn/ONCE}, LKPNR\footnote{https://github.com/Xuan-ZW/LKPNR}) on the MIND dataset.  
Specifically, we only reproduce DIRE (Discriminative Recommendation Framework)-NAML to conduct our main experiments in Table \ref{tab:4}.
Each model is trained and tested five times.
For the sake of fairness and impartiality of the results, we calculate their average and standard deviation as shown in Table \ref{tab:4}, Table \ref{tab:div_cate_adressa} and Table \ref{tab:div_cate_mind}.

\subsection{Performance Evaluation}

\subsubsection{Reply to Q1: LLM-based news recommendation models vs. deep learning-based news recommendation models \textit{w.r.t.} classification and ranking}

As we can see in Table \ref{tab:4}, the classification metrics including AUC and Recall reflect the performance of positive and negative classification in news recommender systems.
For experiments on the MIND dataset, the LLM-based news recommendation model MANNeR achieves the best performance in terms of the main metric AUC, improving \footnote{The improvement is over the suboptimal performing baseline methods.} by approximately 13\% compared to the deep learning-based news recommendation model TANR.
Moreover, LKPNR significantly outperforms the best deep learning-based news recommendation model NAML in terms of Recall metrics, achieving an average 7\% improvement.
These improvements prove the effectiveness of LLM-based news recommendation models in terms of classification and ranking capability on the MIND dataset.
However, results on the Adressa dataset exhibit conflicting phenomena. 
Notably, deep learning-based news recommendation models such as CAUM and LSTUR markedly outperform all LLM-based news recommendation models.
This happens to be because it isn't effective for LLM-based news recommendation models to encode multilingual information such as Norwegian on the Adressa dataset. 


\begin{figure*}[htbp]
  \centering  \includegraphics[width=0.95\textwidth]{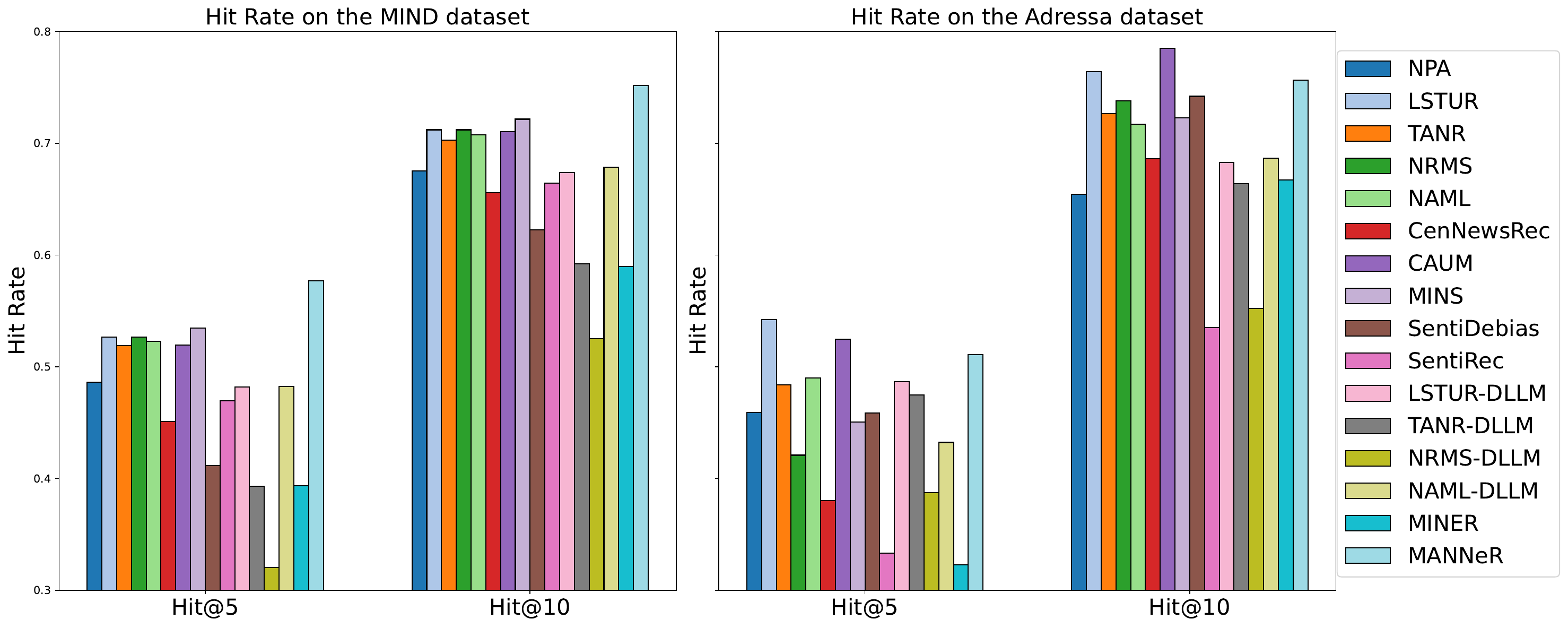} 
  \caption{The Hit Rate on the MIND and Adressa dataset.}
  \label{fig:MIND_Hit_Scores}
\end{figure*}


To more intuitively evaluate the accuracy of news recommendation models, we enrich our experiments using Hit Rate and Precision. 
Figure \ref{fig:MIND_Hit_Scores} demonstrates the results in terms of Hit Rate and Precision on the MIND dataset.
MANNeR achieves the best performance \textit{w.r.t.} Hit Rate on the MIND dataset.
This superiority can be attributed to the effectiveness of a novel aspect-based framework with LLM-based methodologies which can learn aspect-specific representations for news recommendation.
In addition, MANNeR obviously outperforms other news recommendation models in terms of Precision as shown in Figure \ref{fig:MIND_Precision_Scores}.
This emphasizes the superiority of LLM-based news recommendation models in terms of Hit Rate and Precision on the MIND dataset. 
However, the results illustrate different tendencies on the Adressa dataset.
To be specific, LSTUR and CAUM stand out as two of deep learning-based news recommendation models as indicated by Hit@$k$.
This is likely due to the fact that LLM-based news recommendation models cannot effectively encode multilingual news features, leading to decreased performance on the Adressa dataset.
In contrast, LLM-based news recommendation models, such as LSTUR-DLLM and NAML-DLLM, exhibit better precision compared to deep learning-based news recommendation models.
This superiority can be attributed to the fact that LLM-based news recommendation models can model informative news representations, which is effective in deriving higher precision.

In summary, LLM-based news recommendation models have demonstrated great potential, while deep learning methods still have their advantages.

\subsubsection{Reply to Q2: LLM-based news recommendation models vs. deep learning-based news recommendation models \textit{w.r.t.} diversity and personalization}

Table \ref{tab:div_cate_mind} presents results of diversity and personalization on the MIND dataset.
Notably, LLM-based news recommendation models like MINER and NRMS-DLLM outperform deep learning-based news recommendation models in terms of diversity, achieving an average 7\% improvement over SentiRec.
Table \ref{tab:div_cate_adressa} indicates that LLM-based news recommendation models outperform deep learning-based news recommendation models on the Adressa dataset, with improvements up to 6\%.
These results verify the superiority of LLM-based news recommendation models in enhancing the diversity and personalization of news recommendations.
However, deep learning-based news recommendation model like LSTUR exhibits better personalization on the MIND dataset, as indicated by \(categ\_pers@k\).
In contrast, deep learning-based news recommendation models, such as SentiRec and MINS, present a better diversity of news recommendations on the Adressa dataset, as illustrated in Table \ref{tab:div_cate_adressa}.

In summary, the experimental results on the two datasets are complex, and LLM-based news recommendation models do not necessarily outperform deep learning-based methods, while deep learning-based methods still benefit news recommendation systems in terms of diversity and personalization. 

\subsubsection{Reply to Q3: DLLM-empowered news recommendation models vs. original news recommendation models}
\begin{figure*}[htbp]
  \centering
  \includegraphics[width=0.95\textwidth]{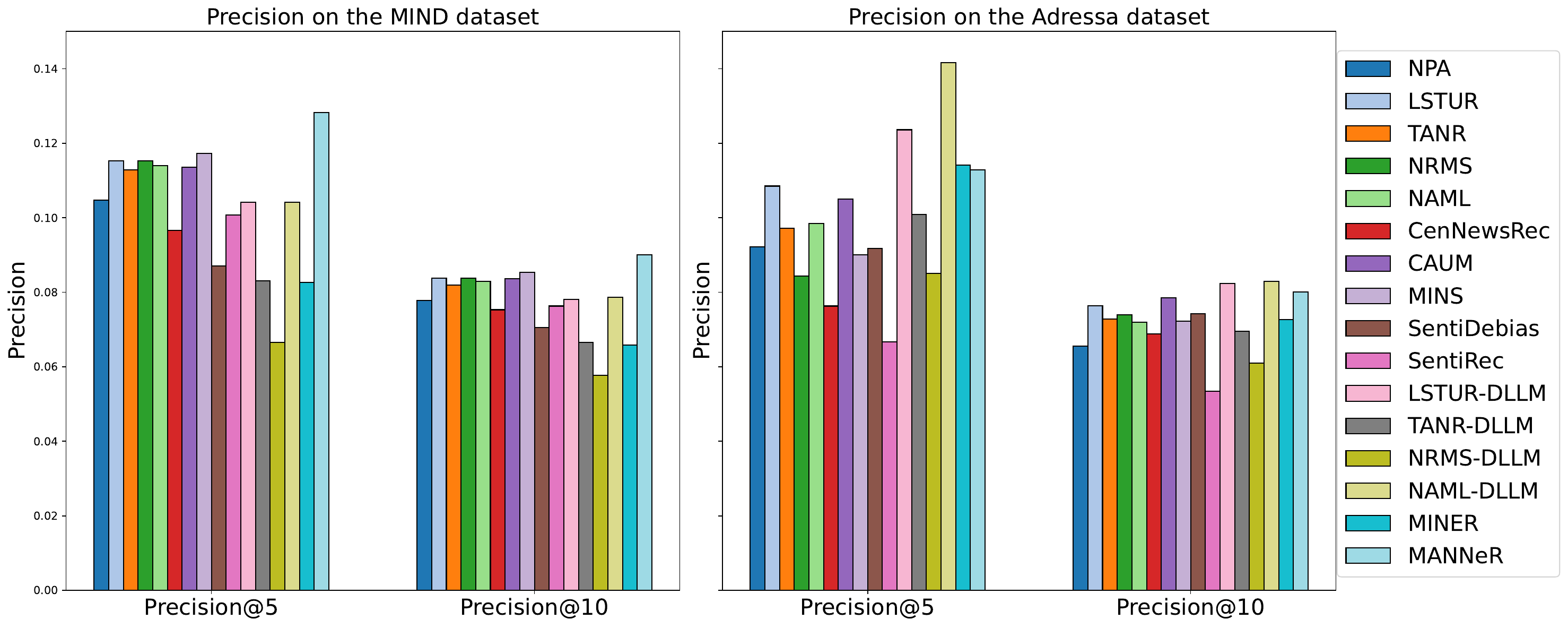} 
  \caption{The Precision on the MIND and Adressa dataset.}
  \label{fig:MIND_Precision_Scores}
\end{figure*}

There are four DLLM-empowered news recommendation models such as LSTUR-DLLM, TANR-DLLM, NRMS-DLLM, and NAML-DLLM in Table \ref{tab:4}.
We can observe that DLLM-empowered news recommendation models do not perform superior to the original models on both datasets.
This may be because DLLMs are not effectively trained on a small-scale dataset.
It is worth noting that a previous study \cite{wu2021empowering} has
demonstrated significant improvements in DLLM-empowered news recommendation models over original models on a large-scale dataset.
In addition, we make an interesting observation: DLLM-empowered news recommendation models show apparent improvements in Precision on the Adressa dataset as illustrated in Figure \ref{fig:MIND_Precision_Scores}.
We assume that DLLM-empowered news recommendation models, such as NAML-DLLM and LSTUR-DLLM, can better capture semantic information from news titles due to the Adressa dataset's emphasis on shallow semantic features, such as titles and keywords.

In addition, we analyze the performance in terms of diversity and personalization as shown in Table \ref{tab:div_cate_mind} and Table \ref{tab:div_cate_adressa}.
We find that most results of DLLM-based news recommendation models illustrate suboptimal performance.
In contrast, NRMS-DLLM achieves the best performance in terms of \(categ\_div@k\) on the MIND dataset, while NAML-DLLM demonstrates significant improvements in terms of \(categ\_pers@k\).

In summary, the improvements of DLLM-empowered news recommendation models are limited by specific factors, such as dataset scale, while deep learning-based news recommendation models continue to offer benefits for news recommendation.

\subsubsection{Reply to Q4: DLLM-based news recommendation models vs. GLLM-based news recommendation models}
As shown in Table \ref{tab:4}, LKPNR exhibits lower performance than MANNeR but still achieves competitive results compared with other baselines.
It is evident that using GLLM-based approaches is not necessarily effective compared with DLLM-based news recommendation models.
Due to the limitation of time and computing resources, we haven't conducted additional experiments on other datasets and metrics.
In the future, we will enrich our experiments and explore more GLLM-based news recommendation models in terms of different metrics.

\begin{table*}[htbp]
  \centering
  \renewcommand{\arraystretch}{1.5}
  \caption{Comparison of different news recommendation models in terms of personalization and diversity on MIND. Through analyzing these results, we can have a conclusion that LLM-based approaches enhance the performance of news recommendation models in terms of diversity and personalization. The reported values represent the mean and standard deviation derived from five distinct experimental runs.
  Bold indicates the best experimental results. Underline means the second-best experimental results.}
    \setlength{\tabcolsep}{2.5pt}
    \begin{tabular}{c|cccccccc}
    \toprule
    \multirow{2}[4]{*}{Model} & \multicolumn{8}{c}{Metric} \\
\cmidrule{2-9}          & categ\_div@5 & categ\_div@10 & categ\_pers@5 & categ\_pers@10 & sent\_div@5 & sent\_div@10 & sent\_pers@5 & sent\_pers@10 \\
    \midrule
    NPA   & 37.05$\pm$0.0075 & 51.09$\pm$0.0073 & 16.96$\pm$0.0019 & 22.69$\pm$0.0019 & 56.67$\pm$0.017 & 67.33$\pm$0.0083 & 25.56$\pm$0.0017 & 34.7$\pm$0.0024 \\
    LSTUR & 29.91$\pm$0.0089 & 43.45$\pm$0.009 & \textbf{20.28$\pm$0.0056} & \textbf{25.69$\pm$0.0043} & 56.28$\pm$0.0179 & 66.13$\pm$0.0087 & 26.25$\pm$0.0024 & \underline{35.11$\pm$0.0021} \\
    TANR  & 35.08$\pm$0.006 & 49.13$\pm$0.0041 & 18.83$\pm$0.0025 & 24.2$\pm$0.0022 & 58.19$\pm$0.0051 & 67.21$\pm$0.0019 & 26.11$\pm$0.0018 & 35.05$\pm$0.0023 \\
    NRMS  & 36.57$\pm$0.0093 & 50.97$\pm$0.0051 & 15.15$\pm$0.0061 & 21.34$\pm$0.0071 & 59.44$\pm$0.0158 & 67.2$\pm$0.0062 & 26.07$\pm$0.0019 & 34.97$\pm$0.0008 \\
    NAML  & 32.74$\pm$0.0046 & 47.4$\pm$0.0046 & \underline{20.06$\pm$0.0027} & 25.09$\pm$0.0026 & 56.93$\pm$0.0066 & 66.77$\pm$0.0041 & 25.92$\pm$0.003 & 34.99$\pm$0.0023 \\
    CenNewsRec & 36.24$\pm$0.008 & 51.04$\pm$0.0034 & 14.74$\pm$0.003 & 20.73$\pm$0.0021 & 59.07$\pm$0.0227 & 67.02$\pm$0.0085 & 25.92$\pm$0.005 & 35$\pm$0.002 \\
    CAUM & 33.67$\pm$0.0073 & 47.17$\pm$0.0026 & 19.28$\pm$0.0032 & \underline{25.21$\pm$0.0033} & 57.78$\pm$0.0125 & 66.56$\pm$0.0077 & 26.21$\pm$0.0022 & 35.2$\pm$0.0011 \\
    MINS  & 32.79$\pm$0.0036 & 46.56$\pm$0.0055 & 19.37$\pm$0.0065 & 24.99$\pm$0.0055 & 56.96$\pm$0.0111 & 66.11$\pm$0.0075 & 26.22$\pm$0.0025 & \textbf{35.16$\pm$0.002} \\
    SentiDebias & 41.11$\pm$0.0082 & 53.58$\pm$0.0059 & 14.69$\pm$0.0059 & 20.97$\pm$0.004 & 57.21$\pm$0.0295 & 66.77$\pm$0.016 & \underline{26.29$\pm$0.0053} & 35.01$\pm$0.0034 \\
    SentiRec & 37.49$\pm$0.0237 & 52.02$\pm$0.0146 & 15.08$\pm$0.0079 & 21.16$\pm$0.0074 & \textbf{60.68$\pm$0.015} & \underline{68.46$\pm$0.0074} & 25.31$\pm$0.0041 & 34.6$\pm$0.0036 \\
    
    \rowcolor{mygray}
    LSTUR-DLLM & 24.68$\pm$0.0232 & 38.19$\pm$0.0174 & 18.88$\pm$0.0074 & 24.06$\pm$0.0064 & 58.23$\pm$0.0298 & 66.97$\pm$0.017 & 25.72$\pm$0.0025 & 34.53$\pm$0.0021 \\
    \rowcolor{mygray}
    TANR-DLLM & 40.17$\pm$0.0177 & 52.09$\pm$0.0167 & 14.03$\pm$0.0135 & 19.7$\pm$0.0165  & 56.46$\pm$0.0231 & 66.03$\pm$0.0161 & 25.13$\pm$0.01 & 33.79$\pm$0.01 \\
    \rowcolor{mygray}
    NRMS-DLLM & \textbf{44.03$\pm$0.0092} & \textbf{54.99$\pm$0.0042} & 14.49$\pm$0.0042 & 20.62$\pm$0.0037 & 59.44$\pm$0.0158 & 67.2$\pm$0.0062 & 26.07$\pm$0.0019 & 34.97$\pm$0.0008 \\
    \rowcolor{mygray}
    NAML-DLLM & 25.87$\pm$0.0268 & 39.38$\pm$0.022 & \textbf{20.46$\pm$0.0079} & 25.07$\pm$0.0078 & 57.28$\pm$0.0149 & 66.48$\pm$0.0084 & 26.09$\pm$0.0039 & 34.89$\pm$0.0038 \\
    \rowcolor{mygray}
    MINER & \underline{42.94$\pm$0.0017} & \underline{54.75$\pm$0.0011} & 15.11$\pm$0.0025 & 21.24$\pm$0.0024 & \underline{60.58$\pm$0.0022} & \textbf{68.71$\pm$0.001} & 25.4$\pm$0.0018 & 34.29$\pm$0.0017 \\
    \rowcolor{mygray}
    MANNeR & 33.95$\pm$0.0147 & 48.06$\pm$0.01 & 19.4$\pm$0.0048 & 24.63$\pm$0.0047 & 51.09$\pm$0.0276 & 63.99$\pm$0.0168 & \textbf{26.32$\pm$0.0018} & 35.1$\pm$0.0005 \\
    \bottomrule
    \end{tabular}%
  \label{tab:div_cate_mind}%
\end{table*}%

\begin{table*}[htbp]
  \centering
  \renewcommand{\arraystretch}{1.5} %
  \caption{Comparison of different news recommendation models in terms of personalization and diversity on Adressa. By comparing deep learning-based news recommendation models with LLM-based news recommendation models in this table, the results demonstrate that general deep-learning methodologies are still effective in processing news data \textit{w.r.t.} diversity and personalization on Adressa. The reported values represent the mean and standard deviation derived from five distinct experimental runs. Bold indicates the best experimental results. Underline means the second-best experimental results.}
  \setlength{\tabcolsep}{2.5pt}
    \begin{tabular}{c|cccccccc}
    \toprule
    \multirow{2}[4]{*}{Model} & \multicolumn{8}{c}{Metric} \\
\cmidrule{2-9}          & categ\_div@5 & categ\_div@10 & categ\_pers@5 & categ\_pers@10 & sent\_div@5 & sent\_div@10 & sent\_pers@5 & sent\_pers@10 \\
    \midrule
    NPA   & 26.13$\pm$0.0112 & 31.34$\pm$0.0025 & 27.25$\pm$0.0019 & 33.78$\pm$0.0028 & 52.48$\pm$0.0064 & 60.72$\pm$0.0041 & 33.09$\pm$0.0012 & 41.92$\pm$0.0007 \\
    LSTUR & 23.13$\pm$0.0107 & 24.87$\pm$0.0111 & \underline{30.67$\pm$0.0124} & 37.49$\pm$0.0098 & 51.91$\pm$0.0062 & 59.87$\pm$0.0037 & 33.18$\pm$0.0012 & \underline{41.99$\pm$0.0007} \\
    TANR  & 25.41$\pm$0.0085 & 30.05$\pm$0.0048 & 28.43$\pm$0.0051 & 34.91$\pm$0.0029 & 52.29$\pm$0.0051 & 60.64$\pm$0.0027 & 33.13$\pm$0.0009 & 41.94$\pm$0.0004 \\
    NRMS  & 27.9$\pm$0.0074 & 31.72$\pm$0.0028 & 26.72$\pm$0.0022 & 34.07$\pm$0.0024 & 52.16$\pm$0.0121 & 60.66$\pm$0.0033 & 33$\pm$0.0011 & 41.88$\pm$0.0004 \\
    NAML  & 26.2$\pm$0.0195 & 30.55$\pm$0.0223 & 29.96$\pm$0.0117 & 36.12$\pm$0.0118 & 52.17$\pm$0.0045 & 60.52$\pm$0.0027 & 33.13$\pm$0.0007 & 41.96$\pm$0.0006 \\
    CenNewsRec & 27.91$\pm$0.0077 & 32.11$\pm$0.0061 & 26.29$\pm$0.0034 & 33.76$\pm$0.0044 & 52.44$\pm$0.0102 & 60.69$\pm$0.0037 & 32.96$\pm$0.0012 & 41.87$\pm$0.0005 \\
    CAUM  & 27.36$\pm$0.0242 & 30.31$\pm$0.0274 & 28.61$\pm$0.0154 & 35.89$\pm$0.011 & 52.61$\pm$0.0073 & 56.83$\pm$0.0836 & \textbf{38.61$\pm$0.124} & 41.93$\pm$0.0005 \\
    MINS  & \underline{30.78$\pm$0.013} & \underline{33.11$\pm$0.0136} & 25.41$\pm$0.0136 & 34.13$\pm$0.0084 & \underline{53.35$\pm$0.0024} & 60.75$\pm$0.0027 & 32.88$\pm$0.0007 & 41.87$\pm$0.0006 \\
    SentiDebias & 27.16$\pm$0.0052 & 31.92$\pm$0.0044 & 26.92$\pm$0.0019 & 33.87$\pm$0.0018 & 52.84$\pm$0.0044 & 60.93$\pm$0.0042 & 33.01$\pm$0.0006 & 41.89$\pm$0.0008 \\
    SentiRec & \textbf{31.05$\pm$0.0027} & \textbf{34.99$\pm$0.0035} & 24.52$\pm$0.0023 & 32.34$\pm$0.0018 & \textbf{57.93$\pm$0.0055} & \textbf{66.98$\pm$0.0035} & 27.32$\pm$0.0035 & 39.17$\pm$0.0011 \\
    \rowcolor{mygray}
    LSTUR-DLLM & 23.43$\pm$0.0416 & 23.45$\pm$0.045 & 28.94$\pm$0.0359 & \underline{38.38$\pm$0.015} & 53.18$\pm$0.0021 & 60.51$\pm$0.0025 & 32.84$\pm$0.0006 & 41.91$\pm$0.0003 \\
    \rowcolor{mygray}
    TANR-DLLM & 19.98$\pm$0.1262 & 22.33$\pm$0.1623 & 29.26$\pm$0.0877 & 35.94$\pm$0.0695 & 51.52$\pm$0.0236 & 59.43$\pm$0.0181 & 32.91$\pm$0.0045 & 41.78$\pm$0.0035 \\
    \rowcolor{mygray}
    NRMS-DLLM & 26.44$\pm$0.0181 & 32.61$\pm$0.016 & 26.62$\pm$0.0096 & 32.77$\pm$0.011 & 52.77$\pm$0.0151 & 61.07$\pm$0.0083 & 32.87$\pm$0.0024 & 41.68$\pm$0.0015 \\
    \rowcolor{mygray}
    NAML-DLLM & 21.65$\pm$0.0451 & 21.9$\pm$0.0696 & \textbf{31.78$\pm$0.0256} & \textbf{39.6$\pm$0.0311} & 52.66$\pm$0.0036 & 60.17$\pm$0.0037 & 33.03$\pm$0.0009 & 41.97$\pm$0.0006 \\
    \rowcolor{mygray}
    MINER & 26.68$\pm$0.0218 & 32.24$\pm$0.0132 & 26.48$\pm$0.0077 & 33.25$\pm$0.0068 & 52.77$\pm$0.0124 & \underline{61.17$\pm$0.005} & 32.92$\pm$0.002 & 41.8$\pm$0.0012 \\
    \rowcolor{mygray}
    MANNeR & 25.56$\pm$0.0123 & 30.91$\pm$0.0138 & 28.36$\pm$0.0064 & 34.44$\pm$0.0056 & 51.69$\pm$0.0139 & 60.06$\pm$0.0088 & \underline{33.28$\pm$0.0023} & \textbf{42.03$\pm$0.0008} \\
    \bottomrule
    \end{tabular}%
  \label{tab:div_cate_adressa}%
\end{table*}%

\section{Future Directions}
\label{section6}
LLM-based news recommendation has already achieved state-of-the-art performance over the past time.
However, there are some challenges to be addressed, including datasets, news-oriented modeling, user-oriented modeling, prediction-oriented modeling, and etc.
In this section, we discuss several promising directions for news recommendation to guide researchers toward more valuable approaches in the future.

\subsubsection{Informative News Recommendation Dataset}
News recommendation dataset is necessary for the development of LLM-based news recommender systems. 
Although several datasets (\textit{e.g.}, MIND, Adressa) are used in news recommendation research, they are insufficient for exploring more comprehensive information such as publisher, image, location, video, and editor.
Due to the lack of this important information, the growth of news recommender systems is limited.
There are three directions to develop LLM-based news recommendation datasets.
Firstly, it is essential to release more informative news recommendation datasets supporting multi-modal and more effective news recommender systems.
Second, as news recommender systems continue to grow, more supervised labels are required.
For example, fairness-aware news recommendation requires bias labels to support this popular direction.
Third, multilingual news recommendation datasets are required to develop multilingual news recommender systems.
Although xMIND \cite{iana2024mind} is a released multilingual news recommendation dataset, it is a translation of the English MIND dataset \cite{wu2020mind} using GLLMs, which compromises the credibility and originality of the source material.

\subsubsection{Deep News Modeling}
News modeling is a basic and critical component in LLM-based news recommender systems.
An effective news encoder is a signal of a superior news recommender system.
This is because clicked news contents represent potential user preferences, which contribute to user modeling. 
The future directions for deep news modeling are as follows.
First of all, multi-modal information is required for deep news modeling.
For example, image and video information could enhance news representations.
Second, multiple aspects of news information need to be applied in news modeling.
For instance, information from publishers may influence fairness in LLM-based news recommender systems.
Third, structural news representations are required to support a deep understanding of news content.
For example, news representations with knowledge graphs could promote understanding of news content.

\subsubsection{Deep User Modeling}
News recommendation focuses on modeling user preferences and then recommending interested news to users.
However, it is difficult to explore accurate and overall user interests.
This is because users' preferences are complex and dynamic in the real-world scenario.
There are three future directions to improve deep user modeling for LLM-based news recommendation.
Firstly, a unified user modeling framework is required in LLM-based news recommender systems.
Most research has explored various types of user preferences such as long- and short-term user interests \cite{An2019LSTUR}, intentions \cite{wang2023intention}, and multiple interests\cite{wang2022news}.
However, there is no unified user modeling framework that supports the representation of various user preferences.
Second, LLM-based news recommender systems are required to understand dynamic user preferences. 
This is because user preferences commonly evolve over time.
For example, breaking news can quickly shift users' interests during that period.

\subsubsection{Effective Prediction Method}
Prediction is a crucial component of news recommender systems, as it determines the outputs of the entire system.
It is common for most research \cite{wu2019neurala, wu2019npa, qi2021hie, wang2022news, lu2022aspect, liu2019hi, li2022miner} to apply MLP and dot-product calculating results in news recommender systems.
As far as we are concerned, two future directions are discussed as follows.
Firstly, prediction with contrastive learning enables to enhance performance of news recommender systems.
This is because contrastive learning excels at maximizing similarity between positive samples while minimizing similarity between negative samples.
Second, prediction with casual inference promotes the interpretability and robustness of news recommender systems.
Due to the black-box nature of deep learning-based methods, the outputs of news recommender systems lack explainability.
Therefore, improving the explainability of predictions has become a crucial direction for future research.

\subsubsection{Efficient Training Strategy}
LLM-based news recommender systems require a large training time to obtain the best performance.
Hence, it is necessary to develop efficient training methods reducing time and resource consumption.
There are two future directions to promote training methods.
First of all, training with knowledge distillation is required in LLM-based news recommender systems.
This is because knowledge distillation helps extract essential information, thereby reducing time and resource consumption.
Second, training with meta-learning can achieve better performance using less data.
Meanwhile, applying meta-learning could help alleviate the cold-start problem in news recommender systems.

\subsubsection{Trustworthy LLM-based News Recommender System}
Building trustworthy news recommender systems is a crucial future direction, encompassing aspects such as modeling fairness, reducing bias, improving explainability, and ensuring privacy preservation in news recommender systems.
It is essential to develop trustworthy news recommender systems due to the incredible outputs of large language models and deep learning-based methods.
Additionally, trustworthy news recommender systems enable to enhance users' experience.
For example, most users prefer to know reasons why systems recommend interesting news articles to them.
Providing explainability can enhance trust in news recommender systems.

\subsubsection{Social-driven LLM-based News Recommender System}
Social-driven news recommender systems can be developed by implementing LLMs.
This is because LLMs can explore related content on the Internet and then learn users' preferences based on real-world relations.
For example, if Tom, who likes the NBA, is shown as Jerry's friend on Facebook, we can infer that Jerry may also like the NBA based on their related online activities.
Therefore, leveraging social relationships on the Internet can help systems extract users' latent preferences and build more accurate user profiles.


\section{Conclusion}
News recommendation is a crucial technology that benefits both individuals and society.
A large amount of previous research demonstrates the superiority of deep learning-based news recommender systems compared to statistical machine learning-based methods. 
With the advancement of large language model technology, news recommendation faces great opportunities and challenges.
In order to assist researchers in studying better this field and guide new researchers in the correct direction, we systematically and comprehensively review LLM-based news recommender systems including methodologies, challenges, and future directions.
Moreover, we propose a unified research framework to review different research.
Notably, it is the first survey to conduct extensive experiments in terms of deep-learning- and LLM-based news recommendation models.
Through this survey, we not only illustrate the superiority of LLM-based news recommender systems but also present the effectiveness of deep learning-based news recommender systems.
In the future, we will evaluate the diversity and personalization of LLM-based news recommendation models and review more LLM-based news recommendation models.

\label{section7}

\section*{Acknowledgments}
This should be a simple paragraph before the References to thank those individuals and institutions who have supported your work on this article.



\section{References Section}
 
\bibliographystyle{IEEEtran}
\bibliography{ref}

\newpage

\vfill

\end{document}